\documentclass[11pt]{article}
\usepackage[letterpaper,margin=1in]{geometry}
\usepackage{amsmath,amssymb,amsthm,mathtools}
\usepackage{microtype}
\usepackage[colorlinks=true,linkcolor=blue,citecolor=blue,urlcolor=blue]{hyperref}
\hypersetup{
  pdftitle={Entropy-Weighted Soundness for Multi-Answer Birthday Graphs},
  pdfauthor={Alexandru Popa},
  pdfsubject={Entropy-weighted biclique counting and unrestricted agreement stability},
  pdfkeywords={birthday repetition, densest k-subgraph, entropy, agreement testing}
}
\newtheorem{theorem}{Theorem}[section]
\newtheorem{lemma}[theorem]{Lemma}
\newtheorem{corollary}[theorem]{Corollary}
\newtheorem{proposition}[theorem]{Proposition}
\newtheorem{remark}[theorem]{Remark}

\usepackage{booktabs}
\usepackage{enumitem}
\usepackage{xcolor}

\newcommand{\val}{\operatorname{val}}

\newcommand{\E}{\mathbb E}
\newcommand{\Prb}{\mathbb P}
\newcommand{\KL}{D_{\mathrm{KL}}}
\newcommand{\bits}{\{0,1\}}
\newcommand{\views}{\binom{[n]}{\ell}}

\title{\bfseries Entropy-Weighted Soundness for Multi-Answer Birthday Graphs}
\author{\textbf{Alexandru Popa}\\[0.4em]
\normalsize Faculty of Mathematics and Computer Science\\
\normalsize University of Bucharest\\
\normalsize Bucharest, Romania}
\date{}
\begin{document}
\pagenumbering{roman}
\maketitle
\thispagestyle{empty}

\begin{abstract}
The vertices of a Boolean birthday graph are partial assignments on
$\ell$-subsets of $n$ variables.  Although a satisfying assignment
induces a $K$-clique for $K=\binom n\ell$, an arbitrary selection of
$K$ vertices may repeat some views and omit others.  This makes
soundness substantially harder than under uniform view sampling.
For bounded-occurrence, constant-gap Boolean ternary CSPs whose
constraints each involve three distinct variables (allowing arbitrary
Boolean predicates), we prove that the graph's maximum
induced $K$-vertex density satisfies
\[
 \rho^*(K)\le 2\exp\!\left(-c\,
 \frac{\ell^3}{n^2\log(en/\ell)}\right)
\]
for $2\le\ell\le x_0n$ and
$\ell\log(en/\ell)\ge C_0\log(n+1)$, with positive constants depending
only on the source parameters.  The argument refines the biclique
counting method by charging the entropy of an adversarially chosen
flattening to mutual information and a combinatorial obstruction
energy.  In the overlapping parameter range it strengthens the explicit
soundness exponent previously established for this graph.

We also establish an unconditional stability theorem for arbitrary sets
$H$ of exactly $K$ distinct local assignments, with missing views and
multiple answers:
$f_*(H)\le34(n/\ell)q_{\mathrm{con}}(H)$ for every
$1\le\ell\le n$.  It gives a complementary additive near-clique
soundness bound.  We do not derive a new fine-grained hardness theorem.
\end{abstract}
\clearpage
\pagenumbering{arabic}

\section{Introduction}\label{sec:introduction}

\paragraph{Why arbitrary selected assignments are difficult.}
A local-assignment graph has a vertex $(S,a)$ for each choice of a small
coordinate set $S$ and a Boolean assignment $a$ on $S$.  A satisfying
constraint-satisfaction assignment induces a large clique: take all its
restrictions.  Soundness asks what an adversary can achieve by selecting the
same number of vertices, but not necessarily one per coordinate set.
The adversary may select many conflicting answers to some views and ignore
other views entirely.  This destroys the uniform view distribution that
underlies elementary birthday-paradox arguments.

For Densest $k$-Subgraph the challenge has two distinct forms.  A selected
subgraph of density $1-o(1)$ ought to encode a single approximately global
assignment.  At substantially lower density no such decoder is guaranteed:
the selection can exhibit many mutually incompatible views, and one instead
needs to count repeated cross-compatibility patterns.  We study both
phenomena for the same Boolean local-assignment graph, with separate
proofs and clearly different ranges of parameters.

\paragraph{Low-density soundness: the main quantitative result.}
Let $\phi$ be a bounded-occurrence Boolean ternary CSP whose
constraints each involve \emph{three distinct variables}, and suppose
every complete assignment violates a constant fraction of its constraints.  Write $n$ for the number of
nonisolated variables, $\ell$ for the view size, $K=\binom n\ell$,
and $x=\ell/n$.  Consider the graph with all
$N=2^\ell K$ local assignments, where two different vertices are adjacent
precisely when they agree on common variables and satisfy every source
constraint whose scope is exposed by the union of their views.  This is the graph of
Manurangsi~\cite{Manurangsi2017rev2026,Manurangsi2026Revision}.

Our first quantitative statement, proved in Section~\ref{sec:legacy-dks-instantiated},
is that for constants $c,x_0,C_0>0$ depending only on the source gap and
occurrence bound, the maximum density $\rho^*(K)$ of any $K$-vertex
induced subgraph satisfies
\begin{equation}\label{eq:intro-cubic}
 \boxed{\rho^*(K)\le 2\exp\!\left(-c\,
 \frac{\ell^3}{n^2\log(en/\ell)}\right)}
\end{equation}
whenever $2\le\ell\le x_0n$ and
$\ell\log(en/\ell)\ge C_0\log(n+1)$.  Its content is especially
visible when the exponent grows, for example when
$\ell^3\gg n^2\log(en/\ell)$.  In comparison, the explicit
soundness theorem in Manurangsi~\cite[Theorem~7]{Manurangsi2026Revision}
proves $\rho^*(K)\le2^{-\delta\ell^4/n^3}$ for
$\ell\ge n^{3/4}/\delta$ and $\ell\le n/2$.
In the overlapping range, the new exponent is larger by a factor of
order $n/(\ell\log(en/\ell))$, up to constants depending on the source
promise.  This comparison concerns the combinatorial density bound;
it does not by itself give a stronger approximation-hardness theorem.

\paragraph{The mechanism: entropy pays for the moving flattening.}
Manurangsi's biclique proof flattens each side of a labelled
$K_{r,r}$ into the set of signed variable values appearing on that
side.  Once these two sets are fixed, three elementary events suppress the
number of compatible tuples: coordinates observed on too few sides;
coordinates assigned both values on one side; and source constraints exposed
by otherwise well-behaved coordinates.  Counting \emph{all} possible
flattenings independently costs as much as $\exp(O(n))$.  The central
observation in our proof is that the entropy of the \emph{random}
flattening, conditioned on biclique acceptance, can be charged to the
same obstruction energy that suppresses a fixed cell, together with the
mutual information between the two biclique sides.  The resulting entropy
bill is $O(\log(e/x))$ per unit of energy, rather than $O(n)$ for a
blind union bound.  This permits $r=\Theta(\log(e/x)/x)$, replacing the
classical $\Theta(1/x^2)$ scale.  The resulting low-density estimate
becomes nontrivial at view sizes of order
$n^{2/3}(\log n)^{1/3}$ (up to source-dependent constants), rather
than the $n^{3/4}$ scale of the earlier explicit bound.  This is a
combinatorial improvement for a specified graph, not an additional
computational hardness theorem.

\paragraph{Agreement stability: a complementary structural theorem.}
Let $H$ be an arbitrary \emph{set} of exactly $K=\binom n\ell$
distinct Boolean local assignments.  Views may be missing or may carry
several answers.  Sample $V,W$ independently, uniformly from $H$, with
replacement.  Let $q_{\mathrm{con}}(H)$ be the probability that these
vertices disagree on a common coordinate.  Let $f_*(H)$ be the fraction
of all $K$ possible views on which the best complete assignment has no
selected restriction.  We prove for every $1\le\ell\le n$ that
\begin{equation}\label{eq:intro-agreement}
 \boxed{f_*(H)\le34\frac n\ell q_{\mathrm{con}}(H).}
\end{equation}
This bound is order-optimal in $n/\ell$ and gives a missing-view
supersaturation principle.  It yields a separate additive near-clique
soundness result for $2\le\ell\le\sqrt n$ with an explicit constant.
The proof uses majority decoding on high-incidence coordinates and a
relative-entropy budget for low-incidence coordinates.

\paragraph{Scope of the results.}
Both theorems handle adversarial multiplicities, but their conclusions
are different.  Agreement stability decodes a global assignment from a
near-consistent selection; entropy-weighted biclique counting bounds
lower-density subgraphs without producing such a decoder.  Neither
result is used to prove the other.  The birthday graph itself and the
fixed-flattening counting framework are due to
Manurangsi~\cite{Manurangsi2017rev2026}; our quantitative refinement is
the entropy charge for a flattening chosen by the selected biclique.
In particular, no new running-time/approximation trade-off under
Almost-ETH follows from the arguments here.

\paragraph{Roadmap.}
Section~\ref{sec:def} specifies the sampling model and the exact
statements.  Section~\ref{sec:method-overview} explains the biclique
proof before any entropy calculation.  Section~\ref{sec:legacy-dks-instantiated}
contains the complete low-density proof.  Section~\ref{sec:related}
compares the nearest prior techniques and parameter regimes.
Appendix~\ref{sec:mainproof} proves unrestricted agreement stability;
Appendix~\ref{sec:dks} gives the additive CSP soundness and source
normalization arguments.  All stated theorems have proofs in the paper.

\section{Models and theorem statements}\label{sec:def}
\paragraph{Birthday graph and the source promise.}
Let $\phi$ be a Boolean CSP on $n$ nonisolated variables with
$m$ constraints, each with a scope of exactly three distinct
variables and an arbitrary predicate on the three Boolean values.
Repeated scopes with different predicates are allowed.  Assume
each variable occurs in at most $d$ constraints and that every
global assignment violates at least an $\varepsilon$ fraction of the
constraints, where $d$ and $\varepsilon>0$ are fixed.
Since every variable occurs and all constraints have arity three,
$n\le3m\le dn$.  Let
$G_{\phi,\ell}$ have one vertex $(S,a)$ for each
$S\in\binom{[n]}\ell$ and $a:S\to\{0,1\}$.
Distinct vertices $(S,a),(T,b)$ are adjacent exactly if they agree on
$S\cap T$ and their union assignment satisfies every constraint of
$\phi$ whose scope lies in $S\cup T$.  The graph has
$N=2^\ell\binom n\ell$ vertices and our target is $K=\binom n\ell$.
An induced $K$-vertex subgraph may contain multiple answers to one
view.  Write $\rho^*(K)$ for its maximum edge density.

\begin{theorem}[Entropy-weighted low-density soundness]\label{thm:lowdens-main}
For every fixed $d\ge1$ and $\varepsilon>0$ there exist
$C_0>0$, $c>0$, and $x_0\in(0,1/2)$, depending only on
$(d,\varepsilon)$, such that for every CSP instance $\phi$ as above
and every integer $\ell$ satisfying
\[
 2\le\ell\le x_0n,\qquad
 \ell\log(en/\ell)\ge C_0\log(n+1),
\]
the birthday graph obeys
\[
 \rho^*(K)\le
 2\exp\!\left(-c\frac{\ell^3}{n^2\log(en/\ell)}\right).
\]
A satisfiable CSP instance yields a $K$-clique.  The graph construction
is the one used in~\cite{Manurangsi2017rev2026}; no complexity-theoretic
inapproximability statement is included in this theorem.
\end{theorem}

\begin{corollary}[The first nonvacuous birthday scale]
\label{cor:scale-threshold}
Fix $d,\varepsilon$ as in Theorem~\ref{thm:lowdens-main} and any
$\theta\in(0,1)$.  There are constants $D_\theta,n_\theta>0$
depending only on $d,\varepsilon,\theta$ such that, whenever
$n\ge n_\theta$ and
\[
 D_\theta n^{2/3}(\log n)^{1/3}\le \ell\le x_0n,
\]
the same graph satisfies $\rho^*(K)\le\theta$.
\end{corollary}
\begin{proof}
For $n$ sufficiently large the displayed lower bound ensures
$\ell\log(en/\ell)\ge C_0\log(n+1)$.  Moreover,
$\log(en/\ell)\le\log(en)$ and hence
\[
 \frac{\ell^3}{n^2\log(en/\ell)}
 \ge \frac{D_\theta^3\log n}{\log(en)}.
\]
Choose $D_\theta$ large enough that the right side, multiplied
by the constant $c$ in Theorem~\ref{thm:lowdens-main}, is at least
$\log(2/\theta)$, and apply that theorem.
\end{proof}

\paragraph{Unrestricted selected-vertex model.}
For $1\le\ell\le n$ let $K=\binom n\ell$, and let $H$ be a \emph{set}
of exactly $K$ distinct Boolean local assignments $(S,a)$ with
$|S|=\ell$.  Take $V,W$ independently uniformly from $H$, with
replacement, and put
\[
 q_{\rm con}(H)=\Pr[\exists i\in S_V\cap S_W:
                       a_V(i)\ne a_W(i)].
\]
For $g\in\{0,1\}^{[n]}$, define
\[
 A_g(H)=\bigl|\{S\in\tbinom{[n]}\ell:(S,g|_S)\in H\}\bigr|,
 \qquad f_*(H)=1-\frac{\max_g A_g(H)}K.
\]
Let $\tau(H)$ be the fraction of the $K$ views having no selected answer.
\begin{theorem}[Unrestricted multi-answer agreement stability]\label{thm:main}
For every $n\ge1$, $1\le\ell\le n$ and selected family $H$ as above,
\[
 f_*(H)\le34\frac n\ell q_{\rm con}(H).
\]
\end{theorem}
\begin{corollary}[Missing-view supersaturation]\label{cor:tau}
For every such $H$,
\[
 q_{\rm con}(H)\ge\frac\ell{34n}\tau(H).
\]
\end{corollary}
\begin{proof}
Since $\tau(H)\le f_*(H)$, apply Theorem~\ref{thm:main}.
\end{proof}

\paragraph{Additive soundness and the boundary between the results.}
Under the stronger promise that every global assignment violates at
least $\gamma n$ constraints with pairwise variable-disjoint three-variable
scopes, Appendix~\ref{sec:dks} proves
\[
 1-\rho^*(K)\ge\frac{\gamma^2}{13600}\frac{\ell^3}{n^2},
 \quad 2\le\ell\le\sqrt n,\ n\ge16/\gamma.
\]
The smaller-view agreement estimate does not prove
Theorem~\ref{thm:lowdens-main}; conversely the biclique theorem
makes no global decoding assertion for an arbitrary selected family.

\section{Proof guide: why the flattening entropy is affordable}\label{sec:method-overview}
\paragraph{Why the previous union bound loses a parameter.}
Let $H$ be any selected $K$-vertex set.  A labelled biclique consists of
$r$ (possibly repeated) left vertices and $r$ right vertices with all
$r^2$ cross edges present.  Write $p_r$ for the probability that $2r$
independent uniform vertices from $H$ form such a tuple.  Alon's
biclique counting inequality gives
\[
 p_r\ge (\rho(H)/2)^{r^2}.
\]
The difficult direction is to upper-bound $p_r$ even when different
selected vertices project onto the same coordinate view.

\paragraph{A fixed flattening can be counted cheaply.}
Replace each vertex by its $\ell$ signed variable-values, and flatten
the entire left and right tuple into signed sets $A,B$.  Every variable
is either \emph{good} (one value occurs on both sides), \emph{bad}
(both bits occur on exactly one side), or \emph{terrible} (at most
one signed occurrence in total).  Write $T$ for the number of terrible
variables and $b$ for the number of bad variables.  If $T$ is large,
the pool of possible vertices has small support.  If $b$ is large,
valid local assignments must avoid many disjoint pairs of opposite
signed values.  If both are small, almost every variable is good,
and many source constraints violated by a completion yield forbidden
signed pairs.  Accordingly, define an energy
\[
  R(A,B)=\max\{T,xb,nx\,\mathbf1_{\{T<\beta n,\ b<\beta n\}}\}.
\]
The fixed-flattening counting lemma states that each cell of accepted
bicliques has size at most $K^{2r}\exp[-c_0rxR(A,B)]$.

\paragraph{The flattening is chosen by the adversary.}
A direct union bound over all signed pairs $(A,B)$ loses
$\exp(O(n))$.  Instead, draw an accepted biclique and regard
its flattening $F=(A,B)$ as a random variable.  The key inequality is
\[
 H(F)\le 5 I(X_L;X_R)+C\log(e/x)\,\mathbb E R(F)+O(\log n).
\]
Its proof encodes bad-role sets relative to the opposite side's
uncovered coordinates; location entropy is charged either to terrible
variables or to $x$ times the number of bad variables.  After
these sets are fixed, the remaining common bits form a shared
function of both biclique sides and cost mutual information.

\paragraph{The density exponent.}
Entropy-weighted cell counting now yields
\[
 -\log p_r\ge(c_0rx-C\log(e/x))\,\mathbb ER(F)
                 -5(-\log p_r)-O(\log n).
\]
With $r=\lceil C_*\log(e/x)/x\rceil$ sufficiently large,
$R(A,B)\ge c_\beta nx$ implies
$-\log p_r=\Omega(nx\log(e/x))$.  Alon's bound,
divided by $r^2$, gives the exponent $nx^3/\log(e/x)$.
Section~\ref{sec:legacy-dks-instantiated} supplies every estimate.

\paragraph{The separate agreement decoder.}
For the near-complete regime, majority-decoding bits at coordinates
with selected incidence at least $\ell/(8n)$ leaves a nearly
consistent family.  Distinctness and $|H|=\binom n\ell$ yield the exact
relative-entropy identity
\[
 D_{\mathrm{KL}}(\mu\Vert u)
 =\log\frac K{(1-\eta)K}+H(A\mid S),
\]
where $\mu$ is the surviving view law and $u$ is the uniform view law.
The entropy cost of rare coordinates forces a global decoder with
error at most $34(n/\ell)q_{\rm con}$.  The complete proof,
including boundary cases, is in Appendix~\ref{sec:mainproof}.

\section{Entropy-weighted biclique counting}
\label{sec:legacy-dks-instantiated}

We use the bounded-occurrence Boolean birthday graph of
Manurangsi~\cite{Manurangsi2017rev2026}.  The same proof also
applies to arbitrary Boolean predicates on ternary scopes, which we
allow throughout this section.  Let $\phi$ be a Boolean CSP on $n$
nonisolated variables, with $m$ constraints involving exactly three
distinct variables each, $\val(\phi)\le1-\varepsilon$, and at most
$d=O(1)$ constraints incident to each variable.  Then $m=\Theta(n)$
and every assignment violates at least $\gamma n$ constraints for
$\gamma=\gamma(\varepsilon,d)>0$.  The only predicate-specific step
below uses this fact: when a global assignment violates a ternary
constraint and two of its three signed variable-values appear in a
left vertex while the third appears in a right vertex, the exposed
constraint makes that pair a nonedge.  The proof does not use that
the predicate is a disjunction.  For CNF sources with shorter clauses,
the fresh-variable, satisfiability- and gap-preserving gadgets of
Lemma~\ref{lem:normal} apply; repeating a literal does not create three
distinct variables.

For $2\le\ell\le x_0n\le n/2$, a vertex of $G_{\phi,\ell}$ is a Boolean partial
assignment whose domain is an $\ell$-subset of the source variables.  We
identify it with an $\ell$-subset of the signed universe
\[
 \mathcal A_\phi=[n]\times\{0,1\}
\]
containing at most one of $(v,0),(v,1)$ for every source variable $v$.
Two vertices are adjacent if their assignments agree on the intersection of
their domains and every source constraint contained in the union of the
two domains is satisfied.  In the YES case, the restrictions of one satisfying
assignment form a clique of size
\[
 K=\binom n\ell.
\]
We write $x=\ell/n$ and $A_0=\log(e/x)$.

The NO-case problem is to bound the density of an \emph{arbitrary} selected
set of exactly $K$ graph vertices; several selected labels may lie above the
same $\ell$-subset.  We use two standard ingredients in exactly the forms
appearing in Manurangsi~\cite{Manurangsi2017rev2026}.
\begin{itemize}[leftmargin=2em]
\item \emph{Labelled bicliques.}  If a graph has edge density $\rho$, then
      the fraction $p_r$ of ordered labelled $K_{r,r}$'s is at least
      $(\rho/2)^{r^2}$~\cite{Alon02}.
\item \emph{Pair avoidance.}  Let $\mathcal P$ be a family of forbidden
      pairs on a universe $U$, and suppose each point of $U$ belongs to at
      most $q$ pairs.  A uniformly random $\ell$-subset of $U$ avoids every
      pair in $\mathcal P$ with probability at most
      \[
       \exp\!\left(-\frac{|\mathcal P|\ell^2}{4q|U|^2}\right)
      \qquad (2\le\ell\le|U|,\ q\ge1).
      \]
\end{itemize}
The first converts biclique information back to DkS density; the second is
used twice inside a fixed flattening.  Everything responsible for the
improved exponent is proved below.  The restriction $\ell\ge2$ in the
second estimate is necessary: a singleton sample avoids every forbidden
pair with probability one.

\begin{lemma}[Pair avoidance under fixed-size sampling]
\label{lem:combined-pairavoid}
Let $\mathcal P$ be a nonempty family of unordered pairs from a ground set
$U$ of size $m$, with maximum pair-incidence degree at most $q$.
If $S$ is a uniformly random $\ell$-subset of $U$, with $2\le\ell\le m$,
then
\[
 \Pr[S\text{ contains no member of }\mathcal P]
 \le\exp\!\left(-\frac{|\mathcal P|\ell^2}{4qm^2}\right).
\]
\end{lemma}
\begin{proof}
Greedy matching extraction leaves a pairwise vertex-disjoint subfamily of
size $t\ge|\mathcal P|/(2q)$: every chosen pair eliminates at most
$2q-1$ remaining pairs.  Under sampling without replacement, the inclusion
indicators of individual points of $U$ are negatively associated
\cite{JoagDevProschan1983}.  Since our $t$ pairs are disjoint,
the decreasing indicators that each pair is \emph{not} fully contained in
$S$ are also negatively associated.  Therefore the probability of avoiding
all matching pairs is at most
\[
 \prod_{j=1}^{t}\left(1-\frac{\ell(\ell-1)}{m(m-1)}\right)
 \le\exp\!\left(-t\frac{\ell(\ell-1)}{m(m-1)}\right)
 \le\exp\!\left(-\frac{|\mathcal P|\ell^2}{4qm^2}\right).
\]
Here the last inequality uses $\ell-1\ge\ell/2$ and $m-1\le m$.
Avoiding all pairs in $\mathcal P$ is stronger than avoiding the matching,
which completes the proof.
\end{proof}

\subsection{Proof roadmap}
\label{subsec:entropy-flattening-closure}

Fix an arbitrary selected $K$-vertex subgraph $H$.  The proof has four steps.
\begin{enumerate}[leftmargin=2em]
\item Sample an accepted labelled biclique $K_{r,r}$ from $H$ and flatten
      each side into the signed variables appearing anywhere on that side.
\item For a fixed flattening, quantify the three obstructions already present
      in the classical proof: terrible variables, bad variables, and violated
      source constraints.  We summarize them by one energy $\mathcal R$.
\item Bound the entropy of the \emph{random} flattening by
      $O(I(X_L;X_R))+O(A_0\,\mathbb E\mathcal R)$.  This is the new step and
      replaces the $2^{\Theta(n)}$ union bound.
\item Choose $r=\Theta(A_0/x)$.  The fixed-cell penalty is then a larger
      multiple of $A_0\mathcal R$ than the entropy needed to name the cell,
      forcing a large biclique-information cost.  Alon's lemma converts this
      cost into the desired DkS density bound.
\end{enumerate}

The notation below implements this roadmap.  The entropy accounting can
be summarized as follows.

\begin{center}
\small
\begin{tabular}{@{}p{0.27\linewidth}p{0.32\linewidth}p{0.30\linewidth}@{}}
\toprule
information & why it can vary & how it is paid \\
\midrule
common bits on good variables & shared by both biclique blocks
  & $I(X_L;X_R)$ \\
moving bad-left/right sets & chosen inside the opposite uncovered set
  & $A_0(t_{\rm terr}+x b_{\rm bad})+I(X_L;X_R)$ \\
terrible locations and states & visible on at most one side
  & $A_0\mathcal R$ \\
\bottomrule
\end{tabular}
\end{center}

The table is a guide; Lemma~\ref{lem:v149-flattening-entropy} below
proves these charges simultaneously.

\begin{lemma}[Entropy-weighted cell counting]
\label{lem:v150-entropy-cells}
Let a finite set $\Omega$ be partitioned as
$\Omega=\bigsqcup_f\Omega_f$, and let $Q$ be uniform on $\Omega$.
If $|\Omega_f|\le M_f$ for every cell, then
\[
 \log|\Omega|
 \le H_Q(F)+\mathbb E_Q\log M_F.
\]
Equivalently, if a common ambient count $M$ satisfies
$M_f\le M e^{-s(f)}$, then
\[
 \boxed{
 \log\frac{M}{|\Omega|}
 \ge \mathbb E_Q s(F)-H_Q(F).}
\]
\end{lemma}

\begin{proof}
Since $Q(F=f)=|\Omega_f|/|\Omega|$,
\[
 H_Q(F)
 =\log|\Omega|-\mathbb E_Q\log|\Omega_F|.
\]
Use $|\Omega_F|\le M_F$ and rearrange.
\end{proof}

\begin{corollary}[Energy-versus-entropy principle]
\label{cor:v153-energy-entropy}
Suppose the cells satisfy
$|\Omega_f|\le M\exp(-aR(f))$ for some nonnegative energy $R$, and
suppose under the uniform law on $\Omega$,
\[
 H(F)\le b\,\mathbb E R(F)+\eta.
\]
Then
\[
 \boxed{
 \log\frac{M}{|\Omega|}
 \ge (a-b)\mathbb E R(F)-\eta.}
\]
In particular, once the fixed-cell coefficient $a$ dominates the entropy
coefficient $b$, the number of possible cells no longer enters separately.
\end{corollary}

\begin{proof}
Apply Lemma~\ref{lem:v150-entropy-cells} with
$s(f)=aR(f)$ and substitute the assumed entropy bound.
\end{proof}

\begin{lemma}[Information-charged adaptive cell counting]
\label{lem:adaptive-cell-counting}
Let $V$ be any set of $K$ points, let $\Omega$ be a nonempty subset
of $V^r\times V^r$, and let $Q$ be uniform on $\Omega$.
Write $(X_L,X_R)\sim Q$, $p=|\Omega|/K^{2r}$, and
suppose that a deterministic cell map $F$ and a nonnegative
cell energy $R(F)$ satisfy, for $a,b,h,\kappa\ge0$,
\[
 |\Omega_f|\le K^{2r}e^{-aR(f)},\qquad
 H_Q(F)\le\kappa I_Q(X_L;X_R)+b\,\mathbb E_QR(F)+h.
\]
Then
\begin{equation}\label{eq:adaptive-cell-abstract}
 (1+\kappa)\log(1/p)
 \ge (a-b)\,\mathbb E_QR(F)-h.
\end{equation}
If additionally $a\ge b$ and $R(f)\ge R_0$ in every nonempty cell,
then the right-hand side is at least $(a-b)R_0-h$.
\end{lemma}
\begin{proof}
Lemma~\ref{lem:v150-entropy-cells} implies
$\log(1/p)\ge a\mathbb ER-H_Q(F)$.
Because $Q$ is the uniform law on $\Omega$ and the ambient law is
uniform on $V^{2r}$, the KL chain rule gives
\[
 \log(1/p)=D_{\rm KL}(Q\|P^r\otimes P^r)
 =I_Q(X_L;X_R)+D_{\rm KL}(Q_L\|P^r)
   +D_{\rm KL}(Q_R\|P^r)\ge I_Q(X_L;X_R),
\]
where $P$ is uniform on $V$.  Substitute the entropy assumption and
rearrange.  The energy-floor conclusion follows from $a\ge b$.
\end{proof}

This elementary lemma is a convenient abstraction of the accounting
used below: the cell penalty pays for a randomly chosen cell, while
mutual information arising from conditioning is charged to the
selection event.  We do not claim the lemma itself as a new general
repetition theorem.

Put
\[
 K=\binom n\ell,\qquad x=\ell/n,\qquad A_0=\log(e/x),
\]
and assume throughout that $x\le x_0$, where $x_0>0$ may be chosen
smaller than the source-dependent constant $\beta$ below, and that
$nxA_0\ge C_0\log(n+1)$ for a sufficiently large constant
$C_0=C_0(d,\varepsilon)$.
Let $H$ be an arbitrary selected subgraph of the birthday graph on exactly
$K$ vertices, and let $P$ be uniform on $V(H)$.

For an integer
\begin{equation}\label{eq:v152-r}
 r=\left\lceil C_*\frac{A_0}{x}\right\rceil
\end{equation}
to be fixed below, let $\mathcal K_{r,r}(H)\subseteq V(H)^{2r}$ be the event
that an ordered tuple
\[
 \mathbf X=(L_1,\ldots,L_r,R_1,\ldots,R_r)
\]
has every cross pair $(L_i,R_j)$ adjacent in $H$.  Repetitions within one side
are allowed, as in the standard notion of a labelled biclique.  Put
\[
 p_r=P^{2r}[\mathcal K_{r,r}(H)].
\]
If $p_r=0$, the desired lower bound on $L_r:=\log(1/p_r)$ is immediate.
Hence assume $p_r>0$ and let
\[
 Q=P^{2r}(\,\cdot\mid\mathcal K_{r,r}(H)).
\]
Since $P$ is uniform on the $K$ selected vertices, $Q$ is exactly the uniform
law on accepted labelled bicliques.  Put
\[
 X_L=(L_1,\ldots,L_r),\qquad X_R=(R_1,\ldots,R_r).
\]
Every birthday vertex is viewed as an $\ell$-subset of the signed universe
$\mathcal A_\phi$.  Its two flattenings are
\[
 \mathsf F=(A,B),\qquad
 A=\bigcup_{i=1}^rL_i,\quad B=\bigcup_{j=1}^rR_j.
\]
For signed sets $A,B\subseteq\mathcal A_\phi$, write
$\mathcal K_{r,r}(A,B)$ for the accepted tuples in the full birthday graph
whose left and right flattenings are exactly $A$ and $B$.  Finally,
\[
 L_r:=\log\frac1{p_r}
 =D_{\rm KL}(Q\|P^{2r}).
 \label{eq:v149-Lt}
\]
All logarithms in this subsection are natural.

\subsubsection{Fixed flattenings: three ways to lose mass}

A flattening remembers only which signed assignments occur somewhere on
each side of the biclique.  If $(v,b)\in A$ and $(v,1-b)\in B$, then the cell
$\mathcal K_{r,r}(A,B)$ is empty: some left vertex contains $(v,b)$ and some
right vertex contains $(v,1-b)$, so that cross pair is inconsistent.  Hence
every nonempty cell has exactly the three variable types below.  The table
also records the mechanism used to penalize each type.

\begin{center}
\begin{tabular}{@{}lll@{}}
\toprule
class & signed pattern & source of exponential loss \\
\midrule
good & one common bit appears on both sides & violated source constraints \\
bad-left/right & both bits on one side, absent on the other & pair avoidance \\
terrible & at most one signed copy appears in total & support deficit \\
\bottomrule
\end{tabular}
\end{center}

Formally, for a cross-consistent flattening $(A,B)$ classify a source variable
$v$ as follows.
\begin{itemize}
\item $v$ is \emph{bad-left} if both $(v,0),(v,1)$ belong to $A$.
      Cross consistency then forces neither signed copy of $v$ to belong to
      $B$.  Define bad-right symmetrically, and write
      \[
       B_L,B_R,\qquad b_{\rm bad}:=|B_L|+|B_R|.
      \]
\item $v$ is \emph{good} if for some bit $b$ one has
      $(v,b)\in A\cap B$.  Cross consistency then forbids $(v,1-b)$ from
      either flattening.
\item $v$ is \emph{terrible} otherwise.  Thus across $A$ and $B$ together
      at most one signed copy of $v$ occurs.  Let $\mathsf T$ be the terrible
      set and put $t_{\rm terr}:=|\mathsf T|$.
\end{itemize}
These classes partition the variables.

Fix $\beta>0$ sufficiently small that $2d\beta<\gamma/2$, and define
the flattening energy
\begin{equation}
 \mathcal R(A,B)
 :=
 \max\left\{
   t_{\rm terr},\,
   x b_{\rm bad},\,
   nx\cdot
   \mathbf1_{\{t_{\rm terr}<\beta n,\ b_{\rm bad}<\beta n\}}
 \right\}.
 \label{eq:v149-energy}
\end{equation}
The normalization is chosen so that the three cases below have the same
form.  Terrible variables give $\exp(-\Theta(rx t_{\rm terr}))$; bad variables give
$\exp(-\Theta(rx\cdot xb_{\rm bad}))$; and the source-gap branch gives
$\exp(-\Theta(rx\cdot nx))$.  Thus all three losses are
$\exp(-\Theta(rx\mathcal R))$.

\begin{lemma}[Quantitative fixed-flattening energy bound]
\label{lem:v149-cell-energy}
There is a constant $c_0>0$, depending only on the source gap and occurrence
bound, such that every cross-consistent flattening $(A,B)$ satisfies
\begin{equation}
 \boxed{
 |\mathcal K_{r,r}(A,B)\cap V(H)^{2r}|
 \le
 K^{2r}
 \exp\!\left(-c_0rx\,\mathcal R(A,B)\right).}
 \label{eq:v149-cell-energy}
\end{equation}
\end{lemma}

\begin{proof}
It is enough to prove the bound in the full birthday graph.

First consider terrible variables.  Since every nonterrible variable
contributes at most two signed points to $|A|+|B|$, while a terrible
variable contributes at most one,
\[
 |A|+|B|\le2n-t_{\rm terr}.
\]
A valid left vertex is an injective $\ell$-subset of $A\subseteq\mathcal A_\phi$.
For an upper bound we may forget injectivity, so there are at most
$\binom{|A|}{\ell}$ possible left vertices, and similarly at most
$\binom{|B|}{\ell}$ right vertices.  For a nonempty cell, both $|A|$ and $|B|$ are at least $\ell$.
Applying AM--GM separately to the $\ell$ factors of the two binomial
coefficients gives the \emph{integer-exact} estimate
\begin{align*}
 \frac{\binom{|A|}{\ell}\binom{|B|}{\ell}}{K^2}
 &=\prod_{j=0}^{\ell-1}
   \frac{(|A|-j)(|B|-j)}{(n-j)^2}\\
 &\le\prod_{j=0}^{\ell-1}
 \left(1-\frac{t_{\rm terr}}{2(n-j)}\right)^2
 \le\exp\!\left(-t_{\rm terr}\sum_{j=0}^{\ell-1}\frac1{n-j}\right)
 \le e^{-xt_{\rm terr}}.
\end{align*}
Here all product factors are nonnegative, and the last two inequalities
follow from $\log(1-z)\le-z$ and $\sum_j1/(n-j)\ge\ell/n$.
This bound has no noninteger-binomial convention or rounding loss, and gives
\begin{equation}
 |\mathcal K_{r,r}(A,B)|
 \le
 K^{2r}e^{-rx t_{\rm terr}}.
 \label{eq:v149-terrible-cell}
\end{equation}

Before the remaining cases, observe also that cross consistency implies
$|A|+|B|\le2n$.  The same AM--GM product comparison with
$t_{\rm terr}=0$ proves
\begin{equation}\label{eq:cell-baseline}
 \binom{|A|}{\ell}\binom{|B|}{\ell}\le K^2.
\end{equation}
This \emph{product} baseline is essential: one side may contain more
than $n$ signed points, so neither side's count is separately at most $K$.
An avoidance probability on one side is multiplied by both binomial counts;
then \eqref{eq:cell-baseline} supplies the required $K^{2r}$ baseline.

Next consider bad variables.  One of $B_L,B_R$ has size at least
$b_{\rm bad}/2$; suppose it is $B_L$.  For each $v\in B_L$ consider the
2-set
\[
 P_v:=\{(v,0),(v,1)\}\subseteq\mathcal A_\phi.
\]
The sets $P_v$ are pairwise disjoint.  No valid left birthday vertex can
contain both points of $P_v$, because a local assignment gives only one value
to the source variable $v$.  Thus the relevant forbidden objects are pairs
of \emph{signed points}, not pairs of source variables.  Applying Lemma~\ref{lem:combined-pairavoid} on the ground set
$A\subseteq\mathcal A_\phi$ (whose size is at most $2n$) shows that a uniformly random
$\ell$-subset of $A$ avoids all these pairs with probability at most
\[
 \exp\!\left(-c_2 b_{\rm bad}\frac{\ell^2}{n^2}\right)
 =
 e^{-c_2b_{\rm bad}x^2},
\]
where we used $|A|\le2n$.  Multiplying the $r$ left-side
avoidance factors by the unrestricted signed-subset counts on \emph{both}
sides, and invoking \eqref{eq:cell-baseline}, gives
\begin{equation}
 |\mathcal K_{r,r}(A,B)|
 \le
 K^{2r}e^{-c_3rb_{\rm bad}x^2}.
 \label{eq:v149-bad-cell}
\end{equation}

Finally suppose $t_{\rm terr}<\beta n$ and $b_{\rm bad}<\beta n$.  Then at least
$(1-2\beta)n$ variables are good.  Their unique common signed values define
a partial assignment $f$; extend $f$ arbitrarily to all source variables.
The source gap gives at least $\gamma n$ constraints violated by $f$.
At most $d(t_{\rm terr}+b_{\rm bad})<2d\beta n<\gamma n/2$
constraints touch a non-good variable, so at least $\gamma n/2$
violated constraints have all variables good.

For each such constraint with scope $C=\{u,v,w\}$ choose two of its
variables, say $u,v$,
and form the signed pair
\[
 P_C:=\{(u,f(u)),(v,f(v))\}\subseteq\mathcal A_\phi.
\]
Because $u,v,w$ are good, all three signed points
$(u,f(u)),(v,f(v)),(w,f(w))$ occur in both flattenings.  In particular, some
right vertex contains $(w,f(w))$.  If a left vertex contained $P_C$, its
union with that right vertex would expose the whole constraint on $C$
with the violating assignment $f$, contradicting that every cross pair
of the biclique is an edge.  This argument uses only the violated
ternary predicate, not a CNF clause.  Hence every left vertex avoids $P_C$.

Because every source variable occurs in at most $d$ constraints, every
signed point belongs to at most $d$ of the chosen pairs $P_C$.
Moreover an identical pair can arise from at most $d$ constraints.  Discarding duplicates therefore
leaves $\Omega(n)$ distinct forbidden signed pairs of incidence $O(d)$ in
$\mathcal A_\phi$.  The same Lemma~\ref{lem:combined-pairavoid} on the left side,
together with \eqref{eq:cell-baseline} for the unrestricted left and right
choices, gives
\begin{equation}
 |\mathcal K_{r,r}(A,B)|
 \le
 K^{2r}e^{-c_4rnx^2}.
 \label{eq:v149-source-cell}
\end{equation}

Set
\[
 R_1=t_{\rm terr},\qquad R_2=xb_{\rm bad},\qquad
 R_3=nx\,\mathbf1_{\{t_{\rm terr}<\beta n,\,b_{\rm bad}<\beta n\}}.
\]
The three arguments above give
\[
 |\mathcal K_{r,r}(A,B)|
 \le K^{2r}e^{-a_i rxR_i}
 \qquad(i=1,2,3),
\]
where $a_1=1$, $a_2=c_3>0$ from
\eqref{eq:v149-bad-cell}, and $a_3=c_4>0$ from
\eqref{eq:v149-source-cell}; the last inequality is trivial when $R_3=0$.
Thus, for $c_0=\min\{a_1,a_2,a_3\}$,
\[
 |\mathcal K_{r,r}(A,B)|
 \le K^{2r}e^{-c_0rx\max_iR_i}
 =K^{2r}e^{-c_0rx\mathcal R(A,B)},
\]
which is \eqref{eq:v149-cell-energy}.
\end{proof}

Notice that, pointwise,
\begin{equation}
 \boxed{
 \mathcal R(A,B)\ge c_\beta nx.}
 \label{eq:v149-energy-floor}
\end{equation}
Indeed, if the indicator in \eqref{eq:v149-energy} is one this is immediate;
otherwise either $t_{\rm terr}\ge\beta n$ or $b_{\rm bad}\ge\beta n$.

\subsubsection{The new step: encoding a moving flattening}

For a flattening let $U_L$ and $U_R$ be the sets of variables that occur
(unsigned) on the left and right respectively.  Thus
$B_L\subseteq[n]\setminus U_R$ and
$B_R\subseteq[n]\setminus U_L$.

The next inequality is the bookkeeping heart of the argument.  If a bad
set of size $b$ is chosen from an opposite uncovered set of size $m$, then
$u=m-b$ coordinates are uncovered for reasons other than that bad set.  Those
$u$ coordinates are terrible.  Hence the entropy of choosing the bad role is
paid either by the terrible-variable energy $u$ or by the bad-variable energy
$xb$.  The logarithmic factor $A_0$ is exactly the price of not knowing which
of these two mechanisms is dominant.

\begin{lemma}[Role-set encoding inequality]
\label{lem:v149-role-binomial}
There is an absolute constant $C>0$ such that for all
$0\le b\le m\le n$,
\begin{equation}
 \boxed{
 \log\binom mb
 \le
 C A_0\bigl((m-b)+xb\bigr).}
 \label{eq:v149-role-binomial}
\end{equation}
\end{lemma}

\begin{proof}
Put $u=m-b$.  If $u=0$ there is nothing to prove.  If $u\ge xb$, then
\[
 \log\binom mb=\log\binom mu
 \le u\log\frac{em}{u}
 \le C A_0u,
\]
because $m/u=1+b/u\le1+1/x$.

If $0<u<xb$, write $z=u/b<x$.  Then
\[
 \log\binom mb
 \le u\log\frac{em}{u}
 =b\,z\log\frac{e(1+z)}z.
\]
For $g(z)=z\log(e(1+z)/z)$,
\[
 g'(z)=\log\frac{1+z}{z}+\frac{z}{1+z}>0.
\]
Hence $g(z)\le g(x)$, and the last expression is at most
$Cbx\log(e/x)=CbxA_0$.
\end{proof}

\begin{lemma}[Entropy of the random flattening]
\label{lem:v149-flattening-entropy}
For the accepted-biclique law $Q$, there is a constant
$C=C(d,\varepsilon)>0$ such that
\begin{equation}
 \boxed{
 H_Q(\mathsf F)
 \le
 5\,I_Q(X_L;X_R)
 +C A_0\,\mathbb E_Q\mathcal R(\mathsf F)
 +O(\log(n+1)).}
 \label{eq:v149-flattening-entropy}
\end{equation}
In particular, since $I_Q(X_L;X_R)\le L_r$,
\begin{equation}
 H_Q(\mathsf F)
 \le
 5L_r
 +C A_0\,\mathbb E_Q\mathcal R(\mathsf F)
 +O(\log n).
 \label{eq:v149-flattening-entropy-L}
\end{equation}
\end{lemma}

\begin{proof}
The proof is an explicit code for the flattening.  We first encode where the
two moving bad roles are; next we encode the terrible variables and their
one-sided states; finally, after those data are fixed, the only remaining
information consists of bits that are common to both biclique blocks.

Let $\mathcal C$ denote all non-good information in the flattening:
the two bad sets $B_L,B_R$, the terrible set, and, for every terrible
variable, which of the five possibilities occurs
(absent from both sides, left with bit $0$, left with bit $1$, right with bit
$0$, or right with bit $1$).
Once $\mathcal C$ is known, the set of good variables is known, and the only
remaining information in the flattening is the common bit on every good
variable.

\emph{Step 1: locate the bad-role sets.}
Since $B_L$ is a function of $X_L$ and $U_R$ is a function of $X_R$,
data processing gives $I(B_L;U_R)\le I(X_L;X_R)$.  Therefore
\[
 H(B_L)
 =I(B_L;U_R)+H(B_L\mid U_R)
 \le I(X_L;X_R)+H(B_L\mid U_R).
\]
Writing $M_R=|[n]\setminus U_R|$ and $b_L=|B_L|$, we have
$B_L\subseteq[n]\setminus U_R$.  Condition first on the random size
$b_L$.  Then
\begin{align*}
 H(B_L\mid U_R)
 &\le H(b_L\mid U_R)+H(B_L\mid U_R,b_L)\\
 &\le \log(n+1)
   +\mathbb E\log\binom{M_R}{b_L}.
\end{align*}
Every variable in $([n]\setminus U_R)\setminus B_L$ is terrible: it
is absent from the right block and, because it is not bad-left, at most one
of its two signed copies appears on the left.  Hence
\[
 M_R-b_L\le t_{\rm terr}.
\]
Lemma~\ref{lem:v149-role-binomial} gives
\[
 H(B_L)
 \le I(X_L;X_R)
   +C A_0\,\mathbb E(t_{\rm terr}+xb_L)
   +O(\log n).
\]
The symmetric estimate for $B_R$ yields
\begin{equation}
 H(B_L,B_R)
 \le
 2I(X_L;X_R)
 +C A_0\,\mathbb E(t_{\rm terr}+x b_{\rm bad})
 +O(\log n).
 \label{eq:v149-bad-entropy}
\end{equation}
Since $\mathcal R\ge t_{\rm terr}$ and $\mathcal R\ge xb_{\rm bad}$ pointwise,
the energy term on the right is already $O(A_0\,\mathbb E\mathcal R)$.

\emph{Step 2: locate and describe the terrible variables.}
Recall that $\mathsf T$ is the terrible set and
$t_{\rm terr}=|\mathsf T|$.  Conditioned on the two bad sets, encode first
$t_{\rm terr}$ and then the set itself.  Formally,
\begin{align*}
 H(\mathsf T\mid B_L,B_R)
 &\le H(t_{\rm terr}\mid B_L,B_R)
   +H(\mathsf T\mid B_L,B_R,t_{\rm terr})\\
 &\le \log(n+1)+\mathbb E\log\binom n{t_{\rm terr}}.
\end{align*}
Pointwise,
\begin{equation}
 \log\binom n{t_{\rm terr}}\le C A_0\,\mathcal R(A,B).
 \label{eq:v149-terrible-location}
\end{equation}
We justify this bound in three exhaustive cases.  If
$t_{\rm terr}\ge nx$, then the binomial-entropy estimate gives
\[
 \log\binom n{t_{\rm terr}}
 \le t_{\rm terr}\log(en/t_{\rm terr})
 \le A_0t_{\rm terr}\le A_0\mathcal R.
\]
If $t_{\rm terr}<nx$ and $t_{\rm terr}\ge\beta n$, then
$\mathcal R\ge t_{\rm terr}\ge\beta n$ and hence
\[
 \log\binom n{t_{\rm terr}}
 \le n\log2\le\frac{\log2}{\beta}\mathcal R.
\]
Finally, if $t_{\rm terr}<\min\{nx,\beta n\}$, then
$\log\binom n{t_{\rm terr}}\le nxA_0$.  When
$b_{\rm bad}<\beta n$, the indicator term in
\eqref{eq:v149-energy} is active and $\mathcal R\ge nx$.
Otherwise $b_{\rm bad}\ge\beta n$ implies
$\mathcal R\ge xb_{\rm bad}\ge\beta nx$.
All cases give \eqref{eq:v149-terrible-location} with
$C=O(1+1/\beta)$.  In particular this constant is allowed to depend
on the source gap and degree bound through the choice of $\beta$.
Given $\mathsf T$, each terrible variable has at most five states, so the
remaining conditional entropy is at most
$\mathbb E[t_{\rm terr}]\log5\le C A_0\,\mathbb E\mathcal R$.
Together with \eqref{eq:v149-bad-entropy}, this proves
\begin{equation}
 H(\mathcal C)
 \le
 2I(X_L;X_R)
 +C A_0\,\mathbb E\mathcal R
 +O(\log n).
 \label{eq:v149-C-entropy}
\end{equation}

\emph{Step 3: encode the common bits on good variables.}
Let $Z$ be the vector of common bits on the good variables.  Once
$\mathcal C$ is fixed, the good-variable set is fixed.  By definition of a
good variable, its unique signed value occurs on both sides, so $Z$ is a
deterministic function of $(X_L,\mathcal C)$ and also of
$(X_R,\mathcal C)$.  The common-function bound therefore gives
\[
 H_Q(Z\mid\mathcal C)
 \le I_Q(X_L;X_R\mid\mathcal C).
\]
The chain rule also gives the exact identity
\[
 I(X_L;X_R\mid\mathcal C)
 =I(X_L;X_R)+I(X_L;\mathcal C\mid X_R)-I(X_L;\mathcal C),
\]
and hence
\[
 I(X_L;X_R\mid\mathcal C)
 \le I(X_L;X_R)+H(\mathcal C).
\]
Thus conditioning on the jointly defined classification can create at most
$H(\mathcal C)$ additional mutual information.

Since $\mathsf F$ is determined by $(\mathcal C,Z)$,
\[
 H(\mathsf F)
 \le H(\mathcal C)+H(Z\mid\mathcal C)
 \le I(X_L;X_R)+2H(\mathcal C).
\]
Insert \eqref{eq:v149-C-entropy}, absorb the lower-order logarithmic term,
and obtain \eqref{eq:v149-flattening-entropy}.

Finally the KL chain rule gives
\[
 L_r
 =I_Q(X_L;X_R)
  +D_{\rm KL}(Q_{X_L}\|P^r)
  +D_{\rm KL}(Q_{X_R}\|P^r),
\]
so $I_Q(X_L;X_R)\le L_r$, proving
\eqref{eq:v149-flattening-entropy-L}.
\end{proof}

\begin{remark}[What the lemma really says]
The constant $5$ is immaterial.  The substantive statement is that every
nat needed to describe a moving flattening is paid either by information
already spent to create the accepted biclique, or by at most an
$O(A_0)$ multiple of the same energy $\mathcal R$ that shrinks a fixed
flattening cell.  This is the step that replaces the raw
$2^{\Theta(n)}$ enumeration.
\end{remark}

\subsubsection{Putting the cell penalty and entropy bound together}

\begin{theorem}[Entropy-weighted biclique soundness]
\label{thm:v149-biclique-soundness}
There are constants $C_*,c>0$, depending only on the bounded-occurrence
source gap, such that for
$r$ chosen as in \eqref{eq:v152-r},
\begin{equation}
 \boxed{
 L_r\ge c\,nxA_0.}
 \label{eq:v149-L-lower}
\end{equation}
\end{theorem}

\begin{proof}
Partition the accepted labelled bicliques according to their flattening.
For a fixed cell, Lemma~\ref{lem:v149-cell-energy} supplies energy
coefficient $a=c_0rx$.  Lemma~\ref{lem:v149-flattening-entropy}, specifically \eqref{eq:v149-flattening-entropy-L}, says that
the cell-name entropy has coefficient $b=CA_0$, plus an information term
$5L_r+O(\log n)$ that is already charged to the accepted-biclique event.
Equivalently, Lemma~\ref{lem:adaptive-cell-counting}, applied with
$\kappa=5$, $a=c_0rx$, $b=CA_0$, and $h=O(\log(n+1))$, gives
\[
 L_r
 \ge
 (c_0rx-CA_0)\mathbb E\mathcal R
 -5L_r-O(\log n).
\]
Therefore
\[
 6L_r
 \ge
 (c_0rx-CA_0)\mathbb E\mathcal R
 -O(\log n).
\]
Since $rx=C_*A_0+O(x)$, choose $C_*=C_*(d,\varepsilon)$
sufficiently large that
$c_0C_*-C\ge c_0C_*/2$.
Here $C$ includes the dependence on $1/\beta$ from the
terrible-location estimate~\eqref{eq:v149-terrible-location}.  By
\eqref{eq:v149-energy-floor},
\[
 L_r\ge c_1A_0(nx)-O(\log n).
\]
The quantitative assumption $nxA_0\ge C_0\log(n+1)$, with $C_0$ large enough to absorb the $O(\log n)$ term, completes the proof.
\end{proof}

\begin{corollary}[Logarithmic-cubic birthday soundness]
\label{cor:v149-log-cubic-soundness}
Let $\rho(H)$ be the density of an arbitrary selected
$K=\binom n\ell$ vertex subgraph of the birthday graph.  Then
\begin{equation}
 \boxed{
 \rho(H)
 \le
 2\exp\!\left(
  -c\,\frac{nx^3}{\log(e/x)}
 \right)}
 \label{eq:v149-density}
\end{equation}
for a constant $c>0$ depending only on the source gap and occurrence bound.
\end{corollary}

\begin{proof}
Alon's labelled-biclique lower bound gives
\[
 p_r\ge(\rho(H)/2)^{r^2}.
\]
Thus
\[
 L_r
 \le
 r^2\log\frac2{\rho(H)}.
\]
Theorem~\ref{thm:v149-biclique-soundness} and
$r=\Theta(A_0/x)$ imply
\[
 \log\frac2{\rho(H)}
 \ge
 \Omega\!\left(
 \frac{nxA_0}{A_0^2/x^2}
 \right)
 =
 \Omega\!\left(
 \frac{nx^3}{A_0}
 \right).
\]
This is the stated bound.  Substituting $x=\ell/n$ and
$A_0=\log(en/\ell)$ into \eqref{eq:v149-density} proves
Theorem~\ref{thm:lowdens-main} for every selected $K$-vertex family
$H$.  In the nonvacuous regime where $nx^3/A_0\to\infty$,
after decreasing $c$ the leading factor $2$ may also be absorbed into
the exponential.
\end{proof}

\section{Comparison with prior work}\label{sec:related}
\paragraph{Manurangsi's original graph and density theorem.}
The entire vertex and edge construction here originates in
Manurangsi~\cite{Manurangsi2017rev2026,Manurangsi2026Revision}.
His Theorem~7 bounds the density by
$2^{-\delta\ell^4/n^3}$ for
$\ell\ge n^{3/4}/\delta$ and $\ell\le n/2$, assuming a constant
source gap and bounded occurrence.  For the same graph and source class
our estimate has exponent $\ell^3/(n^2\log(en/\ell))$.
The fixed-flattening obstruction estimates and the biclique conversion
are inherited from that framework.  Our quantitative distinction is
the entropy-weighted treatment of the random flattening, rather than
a union bound over all its possible values.
The graph is not the ordinary FGLSS graph because the edge test checks
all source constraints exposed by the \emph{union} of two local views.
The extension from Boolean 3-CNF to arbitrary Boolean ternary
predicates does not change the proof: the forbidden-pair extraction
requires only that every violated predicate has a three-variable
scope and bounded incidence.  In the
shared range $\ell/n\to0$ our \emph{formal exponent} improves on
$\ell^4/n^3$ by the ratio
$n/[\ell\log(en/\ell)]$ up to source-dependent constants;
for $\ell=\Theta(n)$ no numerical dominance is claimed without
comparing the constants.  A change of the view size cannot be used to
infer our theorem from Theorem~7 because it also changes the graph,
target $K$, and adversarially selected families.

\paragraph{Information theory and earlier birthday repetition.}
Braverman, Ko, Rubinstein and Weinstein~\cite{BKRW2017} already handle
repeated answers and ignored questions with a question/answer entropy
dichotomy.  In their full arXiv v1, Lemma~4.16 relates answer entropy
to agreement violations and Lemma~4.24 relates sufficiently high
question entropy to exposure of violated source constraints.  Both
conclusions are directed toward soundness in their birthday-repetition
construction.  Neither lemma is stated as a bound on the entropy of
\emph{a two-sided signed flattening} under the law conditioned on an
accepted labelled biclique.  Here the specific estimate is
$H_Q(A,B)\le 5I_Q(X_L;X_R)+O(A_0\mathbb E_Q\mathcal R)$:
common coordinate bits are charged to the induced correlation between
the two biclique sides, and the locations of one-sided roles are charged
to the same obstruction energy that shrinks a fixed cell.
This is a precise difference in the intermediate inequalities, not a
claim that the general information-theoretic ideas of BKRW cannot be
adapted to yield the present result.  Such a derivation would affect
publication priority and remains a point for expert comparison.
Alon's labelled-biclique lemma~\cite{Alon02} is used unchanged.
The pair-avoidance estimate is also classical and its precise
form appears in Manurangsi's paper; we supply a direct self-contained
proof subject to standard negative association.

\paragraph{Agreement testing.}
Classical and recent direct-product and agreement tests
\cite{DinurGoldenberg2008,DinurSteurer2014,DinurFilmusHarsha2025}
assume a prescribed query law or a single response on each view.
Structured list agreement is studied by
Gotlib and Kaufman~\cite{GotlibKaufman2023}.  The present stability
statement samples arbitrary selected \emph{vertices}, allowing
missing views and several answers, while scoring a global decoder on
all $K$ possible views.  This vertex-sampling model differs from the single-answer and
prescribed-query formulations; the result is stated independently
rather than as a general separation from those techniques.

\paragraph{Complexity-theoretic scope.}
Manurangsi's revised preprint~\cite{Manurangsi2026Revision}
formulates its hardness implications using Almost-ETH.  Our theorem is
instead a combinatorial implication from a fixed source-CSP promise.
To obtain a fine-grained hardness result, one would need a
\emph{quantitatively controlled} reduction taking a source instance of
size $s$ to a bounded-occurrence, constant-gap, exact-ternary CSP of
size $n=n(s)$, with a certified construction time $T_{\rm PCP}(s)$.
For each repetition size $\ell$, the birthday graph has
\[
 N=2^\ell\binom n\ell,\qquad K=\binom n\ell,
 \qquad \log N\le \ell\log(2en/\ell),
\]
and a straightforward explicit construction takes
$O(N^2(m+\ell))$ time for $m=O(n)$ source constraints.
One would then have to choose $\ell=\ell(s)$, compare the
completeness/soundness ratio $1/\rho^*(K)$ with the target
approximation guarantee, and show that
$T_{\rm PCP}(s)+O(N^2(m+\ell))+T_{\mathcal A}(N)$
contradicts a \emph{specified} source-time assumption for a hypothetical
algorithm $\mathcal A$.  The exponent improved here is just one input
to this calculation; none of these additional steps is supplied by the
present proof.  In particular we do not assert a new Almost-ETH
hardness-vs-running-time trade-off.

\clearpage
\section{Discussion and open problems}\label{sec:discussion}
Entropy-weighted accounting for a flattening chosen by the sampled
biclique strengthens the density exponent obtained from a fixed-cell
union bound on the Boolean birthday graph.  Our theorem is
combinatorial: obtaining an improved multiplicative Densest
$k$-Subgraph hardness trade-off under Almost-ETH would require
additional fine-grained reduction analysis.

Unrestricted agreement stability provides a different conclusion:
a global decoder for any exactly-$K$ selected family with sufficiently
small coordinate inconsistency.  Although both arguments address
missing views and multiple answers, neither uses the other as a lemma.
The distinction is quantitative.  For a selected family $H$, every
coordinate-inconsistent pair is a nonedge, so
\[
 q_{\rm con}(H)\le (1-\rho(H))\frac{K-1}{K}\le 1-\rho(H).
\]
Consequently the agreement theorem only guarantees a nontrivial
global decoding bound from density alone when
$1-\rho(H)<\ell/(34n)$; at $\ell=\sqrt n$ this requires a
vanishing deficit of order $n^{-1/2}$.  In contrast, the biclique
estimate begins to force density bounded away from one when
$\ell^3/(n^2\log(en/\ell))$ is at least a sufficiently large
constant, at the $n^{2/3}(\log n)^{1/3}$ scale.
Neither estimate currently bridges the intervening range.
The agreement theorem is therefore retained as a self-contained
companion result, not presented as a step in the proof of low-density
soundness.  A natural open problem is an interpolating bound that
controls adversarial selections without presupposing near-consistency
or requiring the cubic birthday-scale exponent to be large.
Further questions concern higher alphabets, sharp constants and
applications that require the strengthened density exponent.

\clearpage
\appendix
\section{Complete proof of unrestricted agreement stability}\label{sec:mainproof}

This section proves Theorem~\ref{thm:main} directly for the \emph{selected-vertex}
sampling experiment of Section~\ref{sec:def}.  We use three laws, which
must not be confused: the uniform law on the original $K$ selected vertices;
the uniform law on a subfamily $G$ obtained by deleting vertices that
violate a partial majority assignment; and the uniform law $u$ on the
$K=\binom n\ell$ \emph{possible views}. All entropies below use natural
logarithms, and $p=\ell/n>0$.

\subsection*{Step 1: decoding coordinates seen frequently}

Write
\[
 z_i=\Prb_{V=(S_V,a_V)\sim H}[i\in S_V].
\]
Since every selected view has size $\ell$, the exact incidence identity is
$\sum_i z_i=\ell$. Call coordinate $i$ \emph{heavy} if $z_i\ge p/8$;
let $L$ be the complementary set of \emph{low-incidence} coordinates.
A heavy coordinate appears in at least one selected vertex because $p>0$.
Choose $g(i)$, for each heavy $i$, as a majority of the selected bits at
coordinate $i$ (ties are broken arbitrarily). Let
\[
 \varepsilon=\Prb_{V\sim H}[\exists\text{ heavy }i\in S_V:
                                     a_V(i)\ne g(i)].
\]

\begin{lemma}[Heavy-coordinate majority decoding]\label{lem:heavy}
For the above partial assignment $g$ and the \emph{with-replacement}
consistency-error probability $q=q_{\rm con}(H)$,
\[
 \boxed{\varepsilon\le\frac{16q}{p}.}
\]
\end{lemma}
\begin{proof}
For every vertex $v$ counted by $\varepsilon$, choose \emph{one}
heavy coordinate $i(v)$ on which its local value differs from $g$.
Conditional on the first sampled vertex being this particular $v$, an
independent $W\sim H$ both contains $i(v)$ and assigns its majority
value with probability at least
\[
 \Prb[i(v)\in S_W]\cdot\frac12
       =\frac{z_{i(v)}}2\ge\frac p{16}.
\]
Every such ordered pair $(v,W)$ is inconsistent.  Hence the law of total
probability, conditioning on the \emph{first} sampled vertex, gives
\[
 q\ge\sum_{v\text{ counted by }\varepsilon}
      \frac1K\cdot\frac p{16}
   =\frac p{16}\varepsilon.
\]
The events from different $v$ are disjoint because they specify distinct
first vertices.  We therefore do not need any independence assumption
between disagreement events at different coordinates. The pair $W=v$
can never be counted: its bit at $i(v)$ is the minority value, not the
majority value.
\end{proof}

If $q\ge p/32$, then $34q/p\ge34/32>1\ge f_*$, so the main theorem is
immediate. Henceforth assume
\begin{equation}\label{eq:step6-lowerror}
 q<p/32,\qquad \varepsilon\le16q/p<1/2.
\end{equation}
The sole purpose of this case distinction is to guarantee that the
surviving family below retains more than half of the selected vertices.
There is no small-error assumption in Theorem~\ref{thm:main} itself.

\subsection*{Step 2: the entropy budget of the surviving family}

Let $G\subseteq H$ consist of the selected vertices agreeing with $g$
on every heavy coordinate in their view, and set
\[
 m=|G|=(1-\varepsilon)K>0.
\]
Sample $(S,A)$ uniformly from $G$; write $\mu$ for its marginal
view law.  Let $u$ be uniform over all $K$ possible $\ell$-views.
Write $\lambda=|L|/n$, and introduce the normalized incidence of low
coordinates \emph{under the surviving-view law},
\[
 x=\frac1\ell\E_{S\sim\mu}|S\cap L|.
\]

\begin{lemma}[Incidence and conditional-entropy budget]
\label{lem:budget}
Under \eqref{eq:step6-lowerror},
\begin{align}
 0\le x&\le\frac{\lambda}{8(1-\varepsilon)}
              \le\frac\lambda4,\label{eq:step6-incidence}\\
 \mathsf H(A\mid S)&\le\ell x\ln2,\label{eq:step6-answerentropy}\\
 \KL(\mu\|u)
 &=\ln\frac Km+\mathsf H(A\mid S)
 \le-\ln(1-\varepsilon)+\ell x\ln2.\label{eq:KLid}
\end{align}
\end{lemma}
\begin{proof}
All $z_i$ refer to the \emph{original} uniform law on $H$. Removing
vertices to form $G$ cannot increase the unnormalized total incidence
on $L$. Thus
\[
 m\,\E_{S\sim\mu}|S\cap L|
 =\sum_{(S,a)\in G}|S\cap L|
 \le\sum_{(S,a)\in H}|S\cap L|
 =K\sum_{i\in L}z_i.
\]
Every $i\in L$ has $z_i<p/8$. Dividing by $m\ell$ gives
\eqref{eq:step6-incidence}; if $L=\varnothing$ this inequality simply
reads $x=\lambda=0$.

For each fixed view $S$, all assignments among the surviving vertices
$(S,a)\in G$ agree with $g$ on $S\setminus L$. Consequently, their
remaining randomness lies in at most $|S\cap L|$ Boolean coordinates;
there are no more than $2^{|S\cap L|}$ possible values. This gives
$\mathsf H(A\mid S=s)\le|s\cap L|\ln2$ on each occupied view, and
averaging over $s\sim\mu$ proves \eqref{eq:step6-answerentropy}.

Finally, the pair $(S,A)$ is uniform on \emph{$m$ distinct pairs}, so
$\mathsf H(S,A)=\ln m$. The entropy chain rule yields
$\mathsf H(S)=\ln m-\mathsf H(A\mid S)$. Since $u(s)=1/K$ for each
possible view, the definition of relative entropy implies
\[
 \KL(\mu\|u)
  =\sum_s\mu(s)\ln\frac{\mu(s)}{1/K}
  =\ln K-\mathsf H(S)
  =\ln(K/m)+\mathsf H(A\mid S).
\]
The convention $0\ln0=0$ handles missing views. Substituting $m/K=1-\varepsilon$
and the entropy bound proves \eqref{eq:KLid}.
\end{proof}

The equality in \eqref{eq:KLid} is the structural use of the fixed total
cardinality and distinctness of local assignments: it would not hold in
this form for a multiset of repeated identical vertices or for an
arbitrary probabilistic weighting of the selected pairs.

\subsection*{Step 3: comparing the surviving views to the uniform slice}

\begin{lemma}[Hypergeometric exponential moments and KL comparison]
\label{lem:slice}
Fix $L\subseteq[n]$ with $\lambda=|L|/n$. Let $u$ be uniform on all
$\ell$-subsets, let $\mu$ be \emph{any} distribution on them, and put
$x=\E_{S\sim\mu}|S\cap L|/\ell$.
For every real $t$,
\begin{equation}\label{eq:mgf}
 \E_{S\sim u}\exp(t|S\cap L|)
 \le(1-\lambda+\lambda e^t)^\ell.
\end{equation}
If $0<\lambda<1$, then
\begin{equation}\label{eq:KLlo}
 \boxed{\KL(\mu\|u)\ge\ell d(x\|\lambda),}
\end{equation}
where $d(x\|\lambda)=x\ln(x/\lambda)+(1-x)\ln((1-x)/(1-\lambda))$
(with the standard continuous convention at $x=0,1$).
\end{lemma}
\begin{proof}
Assign positive weights $w_i=e^t$ for $i\in L$ and $w_i=1$
otherwise. The expectation on the left of \eqref{eq:mgf} equals
\[
 \frac{1}{\binom n\ell}\sum_{|S|=\ell}\prod_{i\in S}w_i,
\]
the $\ell$th elementary symmetric \emph{mean} of $w_1,\ldots,w_n$.
Maclaurin's inequality states that this mean is at most the $\ell$th
power of the arithmetic mean of the weights, which is
$1-\lambda+\lambda e^t$. All $w_i$ are strictly positive for every
finite $t\in\mathbb R$, so this comparison is valid for negative as
well as positive tilts. At $\ell=n$, Maclaurin's inequality reduces
to the ordinary arithmetic--geometric mean inequality.

The finite-space variational inequality for relative entropy says
\[
 \KL(\mu\|u)\ge
 t\E_\mu|S\cap L|-\ln\E_u\exp(t|S\cap L|)
\]
for every real $t$ (it follows by comparing $\mu$ with the probability
law proportional to $u(s)e^{t|s\cap L|}$).
The moment bound thus gives
\[
 \KL(\mu\|u)\ge
 \ell\bigl(tx-\ln(1-\lambda+\lambda e^t)\bigr).
\]
For $0<\lambda<1$, the supremum over $t\in\mathbb R$ is precisely
$\ell d(x\|\lambda)$: for $0<x<1$ choose
$e^t=x(1-\lambda)/(\lambda(1-x))$; for $x=0,1$ take the limit as
$t\to-\infty,+\infty$, respectively. This proves \eqref{eq:KLlo}.
\end{proof}

\begin{lemma}[Separating rare-coordinate bias from answer entropy]
\label{lem:binarykl}
For $0<\lambda<1$ and $0\le x\le\lambda/4$,
\begin{equation}\label{eq:numerical}
 d(x\|\lambda)-x\ln2
 \ge\lambda\left(\frac34-\frac14\ln8\right)
 >\frac{2\lambda}{9}.
\end{equation}
\end{lemma}
\begin{proof}
Using $\ln y\ge1-1/y$ with
$y=(1-x)/(1-\lambda)>0$, we have
\[
 (1-x)\ln\frac{1-x}{1-\lambda}\ge\lambda-x.
\]
Therefore, with $t=x/\lambda\in[0,1/4]$,
\[
 d(x\|\lambda)-x\ln2
 \ge\lambda-x+x\ln(x/\lambda)-x\ln2
 =\lambda\bigl(1-t+t\ln(t/2)\bigr).
\]
The value at $t=0$ is interpreted by continuity. The derivative of
$1-t+t\ln(t/2)$ is $\ln(t/2)<0$ on $(0,1/4]$; its minimum on the
interval is $3/4-(\ln8)/4\approx0.230139>2/9$.
\end{proof}

\begin{lemma}[Low-incidence coordinates have small total support]
\label{lem:entropy}
Under \eqref{eq:step6-lowerror}, with $\lambda=|L|/n$ and the
heavy-coordinate error $\varepsilon$,
\[
 \boxed{\ell\lambda\le9\varepsilon.}
\]
\end{lemma}
\begin{proof}
If $L=\varnothing$, then $\lambda=0$ and the result is immediate.
If $L=[n]$, then all $z_i<p/8$, contradicting
$\sum_i z_i=\ell=np$. Thus $0<\lambda<1$ in the remaining case.
Lemma~\ref{lem:budget} gives $0\le x\le\lambda/4$ and the KL upper
bound \eqref{eq:KLid}; Lemmas~\ref{lem:slice} and
\ref{lem:binarykl} give
\[
 -\ln(1-\varepsilon)
 \ge\ell\bigl(d(x\|\lambda)-x\ln2\bigr)
 >\frac{2\ell\lambda}{9}.
\]
Because $0\le\varepsilon<1/2$, the inequality
$-\ln(1-\varepsilon)\le2\varepsilon$ yields
$\ell\lambda\le9\varepsilon$. This also covers $\varepsilon=0$:
then the displayed strict lower bound rules out any nonempty $L$.
\end{proof}

\subsection*{Step 4: the global assignment and boundary cases}

\begin{proof}[Proof of Theorem~\ref{thm:main}]
For $q\ge p/32$, the bound follows from $f_*\le1$ as observed above.
For $q<p/32$, extend the majority assignment $g$ from heavy coordinates
to the coordinates of $L$ arbitrarily. A selected vertex that agrees
with $g$ on heavy coordinates and whose view avoids $L$ agrees with
$g$ on its entire view. Hence, using the \emph{original} law on $H$,
\[
 f_*(H)
 \le\varepsilon+
    \Prb_{V\sim H}[S_V\cap L\ne\varnothing]
 \le\varepsilon+\sum_{i\in L}z_i.
\]
By the definition of a low-incidence coordinate and
Lemma~\ref{lem:entropy},
\[
 \sum_{i\in L}z_i\le\frac p8|L|
 =\frac{\ell\lambda}{8}
 \le\frac{9\varepsilon}{8}.
\]
Therefore
\[
 f_*(H)\le\frac{17\varepsilon}{8}
 \le\frac{17}{8}\cdot\frac{16q}{p}
 =34\frac n\ell q.
\]
The proof includes $q=0$: Lemma~\ref{lem:heavy} forces
$\varepsilon=0$, the entropy argument forces $L=\varnothing$, and
therefore all $K$ selected vertices agree with a single global
assignment. It includes $\ell=1$, when views are single coordinates,
and $\ell=n$, when $K=1$ and the sole selected assignment extends to a
global assignment. All distributions conditioned on $G$ are used only
when $m=(1-\varepsilon)K>0$.
\end{proof}

\begin{remark}[Order-optimal dependence on $n/\ell$]\label{rem:sharpagreement}
Fix a coordinate $i$ and choose exactly one local assignment for each view. Set all bits other than $i$ to zero, and assign $i$ the value one on exactly $b=\lfloor\mu/2\rfloor$ of the $\mu=\binom{n-1}{\ell-1}$ views containing $i$, assigning it zero on the remaining views. Then
\[
 f_*=\frac bK,\qquad q=\frac{2b(\mu-b)}{K^2}.
\]
When $\mu\ge2$ (so $b>0$ and $q>0$), this yields
\[
 \frac{f_*}{(n/\ell)q}=\frac{\mu}{2(\mu-b)}.
\]
As $\mu\to\infty$ the last expression tends to one. Thus the $n/\ell$ dependence in Theorem~\ref{thm:main} cannot be replaced by $o(n/\ell)$ in general. This example says nothing about the optimal numerical constant $34$.
\end{remark}

\section{Complete additive near-clique soundness proof}\label{sec:dks}

We now apply the agreement theorem to a previously studied graph reduction.
The soundness proof first decodes a near-global assignment from a dense
selected subgraph, then finds a violated ternary constraint exposed by
two views supporting that assignment.  A selected set of $K$ graph
vertices is precisely a family to which
Theorem~\ref{thm:main} applies, even if the set includes different
answers over the same view and leaves other views unrepresented.  The
additional task is to convert a decoded global assignment that violates
many constraints into nonedges in the selected subgraph.  We carry out
this conversion with a robust exposure experiment on two \emph{uniform}
views and then justify, explicitly, its transfer to selected vertices.

\subsection{The graph and perfect completeness}

A Boolean ternary CSP $\Phi$ has variable set $[n]$.  Every constraint
has a scope $C\subseteq[n]$ of size \emph{exactly three} and a Boolean
predicate on those variables.  Constraints with equal scopes are allowed;
the general soundness theorem does not require the predicates to be
3-CNF clauses.  For every $S\in\views$ and $a:S\to\bits$, create a
vertex $(S,a)$.  Consequently
\begin{equation}\label{eq:dks-sizes}
 N=|V(G_{\Phi,\ell})|=2^\ell\binom n\ell,
 \qquad K=\binom n\ell.
\end{equation}
For \emph{distinct} vertices $v=(S,a)$ and $w=(T,b)$, put an edge
between $v,w$ exactly if both conditions hold:
\begin{enumerate}
\item $a(i)=b(i)$ for every $i\in S\cap T$;
\item the unique induced assignment on $S\cup T$ satisfies every
constraint of $\Phi$ whose complete three-variable scope lies in
$S\cup T$.
\end{enumerate}
This is a simple, loopless graph.  In particular, two distinct
assignments over the \emph{same} view are always nonadjacent, and no
pair of labels is tested against a constraint whose scope is not fully
contained in their union.  A vertex whose own assignment violates an
internally exposed constraint is isolated, because its violation remains
present in the union with every other vertex.  It remains in the graph:
deleting locally invalid assignments would change the exact cardinality
and is not part of this reduction.  The vertex and edge definitions are
those of Manurangsi~\cite[Section~3]{Manurangsi2017rev2026}.

\begin{lemma}[Perfect completeness]\label{lem:complete}
If $\Phi$ is satisfiable, $G_{\Phi,\ell}$ contains a clique of size
$K=\binom n\ell$.
\end{lemma}
\begin{proof}
Let $g:[n]\to\bits$ satisfy every constraint of $\Phi$ and select
$H_g=\{(S,g|_S):S\in\views\}$.  This consists of $K$ distinct
vertices.  Any two restrictions agree on their common coordinates,
and their union coincides with $g$ on its domain; hence every
constraint exposed in the union is satisfied.  Every two \emph{distinct}
vertices in $H_g$ are adjacent, as required.
\end{proof}

\paragraph{The soundness hypothesis.}
A Boolean CSP is merely \emph{unsatisfiable} if every global assignment
violates at least one constraint.  That is far too weak for the
quantitative bound below.  Our assumption is that for every
$g:[n]\to\bits$ there is a collection of at least $\gamma n$
constraints \emph{violated by $g$} whose scopes are pairwise
disjoint, with $0<\gamma\le1/3$.  The disjoint collection may depend
on $g$.  In particular, the hypotheses imply that $\Phi$ has at
least $\gamma n$ violated constraints for every $g$, but the converse
without a bounded-occurrence assumption need not hold.  The disjoint
scopes are what give the needed negative correlations.

\subsection{Robust exposure of violated ternary constraints}
\label{subsec:exposure}

Fix an assignment $g$ and choose $t\ge\gamma n$ violated
constraints on pairwise disjoint scopes
$C_j=\{a_j,b_j,c_j\}$, $j\in[t]$.  All $3t$ variables are distinct;
we distinguish a designated variable $c_j$ from the other two solely
to orient the random experiment.  Sample $S,T$ independently and
uniformly from $\views$, \emph{with replacement}.  Set
\begin{equation}\label{eq:exposure-defs}
 \alpha=\frac{\ell(\ell-1)}{n(n-1)},\quad
 \Lambda=t\alpha,\quad p_0=\frac\ell n\Lambda,
 \qquad
 I_j=\mathbf1\{c_j\in S,\ \{a_j,b_j\}\subseteq T\},
 \quad X=\sum_{j=1}^t I_j.
\end{equation}
The event $I_j=1$ ensures $C_j\subseteq S\cup T$; it is only one
of the possible ways the constraint may be exposed, so counting it
provides a lower bound without having to enumerate overlapping
configurations.  Let $B\subseteq\views$ be an arbitrary set of
``bad'' views, with $|B|\le fK$, and put
\begin{equation}\label{eq:exposure-Y}
 Y=X\,\mathbf1\{S\notin B,\ T\notin B\}.
\end{equation}
We seek a positive probability that a constraint is exposed while
\emph{both} sampled views are good.  It would be incorrect to assume
the event $S\notin B,T\notin B$ is independent of $X$; instead, we
bound the contribution of each bad-view event to the first moment.

\begin{lemma}[Robust ternary clause exposure]\label{lem:exposure}
Suppose $2\le\ell\le\sqrt n$ and the violated constraints in
\eqref{eq:exposure-defs} have pairwise disjoint three-variable scopes.
Then
\begin{align}
 \mathbb E X&=p_0,&
 \mathbb E X^2&\le p_0+p_0^2,\label{eq:expmoments}\\
 \mathbb E Y&\ge p_0\left(1-\frac f\gamma
       -\sqrt{\frac{f(1+\Lambda)}{\Lambda}}\right).
 \label{eq:expdeletion}
\end{align}
If $f\le\gamma\Lambda/200$, then
\begin{equation}\label{eq:expconclusion}
 \Prb(Y>0)\ge\frac38p_0
      \ge\frac{3\gamma}{16}\frac{\ell^3}{n^2}.
\end{equation}
\end{lemma}
\begin{proof}
\noindent\textbf{First and second moments.}\ 
For $U$ uniform over $\views$ and a fixed set $A$ of $h$ distinct
coordinates, the exact inclusion probability is
\begin{equation}\label{eq:fixed-size-inclusion}
 \Prb(A\subseteq U)=\frac{(\ell)_h}{(n)_h},
\end{equation}
where $(r)_h=r(r-1)\cdots(r-h+1)$, with the probability interpreted
as zero if $h>\ell$.  As $S,T$ are independent,
\[
 \E I_j=\frac\ell n\frac{(\ell)_2}{(n)_2}
       =\frac\ell n\alpha,
 \qquad \E X=t\E I_j=p_0.
\]
For any two disjoint fixed sets $A,A'$ with $|A|\le\ell$,
conditioning on $A\subseteq U$ can only decrease the probability of
also including $A'$:
\begin{align*}
 \Prb(A'\subseteq U\mid A\subseteq U)
 &=\frac{(\ell-|A|)_{|A'|}}{(n-|A|)_{|A'|}}\\
 &\le\frac{(\ell)_{|A'|}}{(n)_{|A'|}}
 =\Prb(A'\subseteq U).
\end{align*}
The inequality is immediate from the product representation (and
continues to hold if a required set cannot fit).  For $j\ne k$,
$\{c_j\},\{c_k\}$ are disjoint, and the pairs
$\{a_j,b_j\},\{a_k,b_k\}$ are also disjoint.  Independence of $S$
and $T$ therefore gives
$\E[I_j I_k]\le(\E I_j)(\E I_k)$, though the $I_j$'s are not
independent.  Expanding the square and using $I_j^2=I_j$ gives
\[
 \E X^2=\sum_j\E I_j+\sum_{j\ne k}\E I_jI_k
 \le p_0+\sum_{j\ne k}\E I_j\E I_k
 \le p_0+p_0^2.
\]

\noindent\textbf{Deleting a bad first view.}\ 
Condition on $S$ before drawing $T$.  Since each designated
$c_j$ is distinct, there are at most $\ell$ of them in $S$.
Thus
\[
 \E[X\mid S]=\alpha\sum_j\mathbf1\{c_j\in S\}
 \le\alpha\ell.
\]
It follows without any independence assumption on $B$ that
\begin{equation}\label{eq:delete-S}
 \E[X\mathbf1\{S\in B\}]
 \le f\alpha\ell
 =p_0\frac{fn}{t}\le p_0\frac f\gamma.
\end{equation}

\noindent\textbf{Deleting a bad second view.}\ 
Put
$Z(T)=\sum_{j=1}^t\mathbf1\{\{a_j,b_j\}\subseteq T\}$.
The disjoint pairs have the same negative-inclusion correlation,
so $\E Z=\Lambda$ and
$\E Z^2\le\Lambda+\Lambda^2$.  Since $S$ is independent of $T$,
\[
 \E[X\mid T]=\frac\ell n Z(T).
\]
Cauchy--Schwarz, applied \emph{only after conditioning on $T$}, gives
\begin{align}
 \E[X\mathbf1\{T\in B\}]
 &=\frac\ell n\E[Z(T)\mathbf1\{T\in B\}]\notag\\
 &\le\frac\ell n\sqrt{\Prb(T\in B)\E Z^2}\notag\\
 &\le p_0\sqrt{\frac{f(1+\Lambda)}{\Lambda}}.
 \label{eq:delete-T}
\end{align}
As $X\ge0$, pointwise
$Y\ge X-X\mathbf1\{S\in B\}-X\mathbf1\{T\in B\}$;
\eqref{eq:delete-S} and \eqref{eq:delete-T} prove
\eqref{eq:expdeletion}.

\noindent\textbf{Quantitative second-moment conclusion.}\ 
Scope disjointness gives $t\le n/3$.  Because
$\ell(\ell-1)\le n-1$ when $\ell\le\sqrt n$,
\begin{equation}\label{eq:LambdaLo}
 \frac\gamma2\frac{\ell^2}{n}
 \le\gamma\frac{\ell(\ell-1)}{n-1}
 \le\Lambda\le\frac13,
 \qquad p_0=\frac\ell n\Lambda\le\frac13.
\end{equation}
Under $f\le\gamma\Lambda/200$, we have
\[
 \frac f\gamma\le\frac1{600},\qquad
 \sqrt{\frac{f(1+\Lambda)}{\Lambda}}
 \le\sqrt{\frac{(1/3)(4/3)}{200}}<\frac1{20}.
\]
Thus $\E Y\ge(3/4)p_0$.  Since $0\le Y\le X$, the second-moment
inequality (Cauchy--Schwarz applied to $Y\mathbf1\{Y>0\}$) implies
\[
 \Prb(Y>0)\ge\frac{(\E Y)^2}{\E Y^2}
 \ge\frac{(3p_0/4)^2}{p_0+p_0^2}
 \ge\frac{27}{64}p_0\ge\frac38p_0.
\]
Using the first inequality of \eqref{eq:LambdaLo} and
$p_0=(\ell/n)\Lambda$ gives the second inequality in
\eqref{eq:expconclusion}.
\end{proof}

\subsection{From exposed violations to missing graph edges}
\label{subsec:soundness}

We stress a sampling distinction that is easy to miss.  An arbitrary
$K$-vertex subset $H$ need not induce a uniform distribution on its
views.  However, once a global assignment $g$ is decoded, each view
$S$ for which $(S,g|_S)\in H$ identifies \emph{one unique selected
vertex}.  Thus an ordered pair of such good views $(S,T)$ maps
injectively to the ordered pair of selected vertices
$((S,g|_S),(T,g|_T))$.  Both ordered-pair experiments assign
probability $1/K^2$ to an individual pair.  We do not need, and do not
assert, that conditioning the original selected-vertex distribution
on its occupied views makes that law uniform.

\begin{theorem}[Soundness of the ternary compatibility graph]\label{thm:dks}
Suppose every assignment to $\Phi$ violates at least $\gamma n$
constraints with pairwise disjoint three-variable scopes, where
$0<\gamma\le1/3$, $n\ge16/\gamma$, and
$2\le\ell\le\sqrt n$.  Then every $K$-vertex induced subgraph of
$G_{\Phi,\ell}$ has edge density at most
\[
 \boxed{1-\frac{\gamma^2}{13600}\frac{\ell^3}{n^2}}.
\]
This quantification includes selections with missing views, multiple
distinct answers on a single view, and locally invalid vertices.
\end{theorem}
\begin{proof}
Choose an arbitrary $H\subseteq V(G_{\Phi,\ell})$ with $|H|=K$.
Let $M_{\mathrm{bad}}$ count its \emph{unordered distinct} nonedges,
and let $q_{\mathrm{bad}}=2M_{\mathrm{bad}}/K^2$ be the probability
that two independently sampled selected vertices are \emph{distinct} and
nonadjacent.
The diagonal selected-vertex pairs contribute zero to this definition.
Every coordinate disagreement forbids adjacency, whence
$q_{\mathrm{con}}(H)\le q_{\mathrm{bad}}$.

Write $\delta_\gamma=\gamma^2/13600$.  Suppose, toward a
contradiction, that
\begin{equation}\label{eq:small-qbad}
 q_{\mathrm{bad}}<\delta_\gamma\frac{\ell^3}{n^2}.
\end{equation}
Theorem~\ref{thm:main} yields a global assignment $g$ with
$g|_S$ represented on all but $fK$ views, for
$f=f_*(H)\le34(n/\ell)q_{\mathrm{con}}(H)$.  Therefore,
\begin{equation}\label{eq:fExp}
 f\le34\frac n\ell q_{\mathrm{bad}}
 <\frac{\gamma^2\ell^2}{400n}
 \le\frac{\gamma\Lambda}{200},
\end{equation}
where $\Lambda$ refers to a disjoint family of $t\ge\gamma n$
constraints violated by this $g$ and the last inequality uses
\eqref{eq:LambdaLo}.  Define the bad-view set
$B=\{S:(S,g|_S)\notin H\}$, so $|B|=fK$.

Apply Lemma~\ref{lem:exposure} to two uniform independent views.
It gives
\begin{equation}\label{eq:good-exposed}
 \Prb(Y>0)\ge\frac{3\gamma}{16}\frac{\ell^3}{n^2}.
\end{equation}
If $Y>0$ and $S\ne T$, both selected restrictions are present in $H$,
are distinct vertices, and their union violates at least one
constraint.  They are thus nonadjacent, even if their individual
restrictions are consistent on $S\cap T$.

\noindent\textbf{Removing the diagonal.}\ 
When $\ell=2$, the event $Y>0$ requires the three distinct
variables $c_j,a_j,b_j$ to be contained in $S\cup T$; it is
impossible when $S=T$ because that view has only two variables.
For $\ell\ge3$, independent uniform sampling yields
$\Prb(S=T)=1/K$.  By the standard binomial-coefficient inequality
$\binom n\ell\ge(n/\ell)^\ell$ and the assumptions
$\ell\le\sqrt n$, $n\ge16/\gamma$,
\begin{equation}\label{eq:diagonal-bound}
 \frac1K\le\left(\frac\ell n\right)^\ell
 \le\left(\frac\ell n\right)^3
 \le\frac\gamma{16}\frac{\ell^3}{n^2}.
\end{equation}
Consequently, also for $\ell=2$ (with no diagonal subtraction),
\begin{align*}
 q_{\mathrm{bad}}
 &\ge\Prb(Y>0,\ S\ne T)\\
 &\ge\left(\frac{3\gamma}{16}-\frac\gamma{16}\right)
       \frac{\ell^3}{n^2}
 =\frac\gamma8\frac{\ell^3}{n^2}
 >\delta_\gamma\frac{\ell^3}{n^2},
\end{align*}
contradicting \eqref{eq:small-qbad}.  Here each good ordered pair
has weight $1/K^2$ under selected-vertex sampling, as explained
above; no conditioning or measure-change factor is missing.

Finally, for $K\ge2$, let
$\operatorname{den}(H)=|E(H)|/\binom K2$.  Because
$M_{\mathrm{bad}}=\binom K2-|E(H)|$,
\begin{equation}\label{eq:density-conversion}
 q_{\mathrm{bad}}=\frac{2M_{\mathrm{bad}}}{K^2}
 =\frac{K-1}{K}\bigl(1-\operatorname{den}(H)\bigr).
\end{equation}
Thus the missing-edge fraction is at least $q_{\mathrm{bad}}$,
which proves the claimed bound.
\end{proof}

\subsection{A matching-order construction}\label{subsec:sharpness}
\begin{proposition}[Sharpness for Boolean ternary predicates]\label{prop:sharp}
If $n$ is divisible by three, a Boolean ternary CSP satisfies the
hypothesis of Theorem~\ref{thm:dks} with $\gamma=1/3$ and its
compatibility graph has a $K$-vertex induced subgraph whose
missing-edge fraction is at most
\[
 \frac{K}{K-1}\frac83\frac{\ell^3}{n^2}
 \le\frac{16}{3}\frac{\ell^3}{n^2},
 \qquad 2\le\ell\le\sqrt n.
\]
For $\ell=2$, the number of unordered nonedges in the displayed
$K$-vertex subgraph is \emph{exactly} $n(n-2)$.
For all sufficiently large multiples of three, the lower bound of
Theorem~\ref{thm:dks} supplies the matching $\Omega(\ell^3/n^2)$
order throughout the stated repetition regime.
\end{proposition}
\begin{proof}
Partition $[n]$ into $n/3$ disjoint triples.  On every triple
$C$ impose \emph{both} contradictory parity requirements
$\bigoplus_{i\in C}x_i=0$ and
$\bigoplus_{i\in C}x_i=1$.  No assignment can satisfy both, and every
assignment violates exactly one predicate on each triple.  The
$n/3$ violated scopes are pairwise disjoint.

Choose an arbitrary global assignment $g$ and the complete
transversal $H_g=\{(S,g|_S):S\in\views\}$.  Its vertices are
coordinate-consistent.  A distinct pair is nevertheless a nonedge
\emph{if and only if} its union contains one of the prescribed
triples, because a fully exposed triple carries two incompatible
requirements.

For one fixed triple $C$ and independent uniform $S,T$, exact
inclusion--exclusion gives
\begin{equation}\label{eq:sharp-pi}
 \pi_{n,\ell}=\Prb(C\subseteq S\cup T)
 =\sum_{j=0}^3(-1)^j\binom3j
  \left(\frac{\binom{n-j}{\ell}}{\binom n\ell}\right)^2.
\end{equation}
To upper-bound it, assign each of the three elements of $C$ to
one of the two required containments $A\subseteq S$ and
$C\setminus A\subseteq T$.  The union of these eight events covers
$C\subseteq S\cup T$ (an element in both views may be assigned
to either).  For each $A\subseteq C$, independence and the exact
inclusion probabilities yield
\[
 \Prb(A\subseteq S,\ C\setminus A\subseteq T)
 =\frac{(\ell)_{|A|}}{(n)_{|A|}}
  \frac{(\ell)_{3-|A|}}{(n)_{3-|A|}}
 \le\left(\frac\ell n\right)^3.
\]
By a union bound over the eight choices and then over all $n/3$
triples,
\begin{equation}\label{eq:sharp-q}
 q_{\mathrm{bad}}(H_g)
 \le\frac n3\pi_{n,\ell}
 \le\frac83\frac{\ell^3}{n^2}.
\end{equation}
The density conversion \eqref{eq:density-conversion} gives the
stated upper bound.  (Events associated with different triples need
not be disjoint; a union bound is sufficient.)

For $\ell=2$, there is a particularly transparent \emph{exact}
count.  Fix one triple $C$.  An unordered pair of distinct two-sets
exposes $C$ in exactly one of two mutually exclusive ways.
If $|S\cup T|=3$, both views are two-subsets of $C$, yielding
$\binom32=3$ unordered pairs.  If $|S\cup T|=4$, the views are
disjoint; choose the two elements of $C$ forming one view (three
choices), and the fourth element outside $C$ ($n-3$ choices), giving
$3(n-3)$ unordered pairs.  Thus one triple causes
$3+3(n-3)=3(n-2)$ nonedges.  No unordered pair of two-element views
exposes \emph{two} different triples, because $|S\cup T|\le4$
whereas the prescribed triples are disjoint.  Summing over $n/3$
triples gives $M_{\mathrm{bad}}=n(n-2)$, with no double counting.
\end{proof}

\subsection{Source normalization and complexity consequences}
\label{subsec:complexity}

Theorem~\ref{thm:dks} is a combinatorial statement about CSPs
satisfying an explicit promise, not an NP-hardness assertion by
itself.  To obtain an ordinary polynomial-time reduction for
\emph{fixed} $\ell$, we start with the standard perfect-completeness,
bounded-occurrence gap-3SAT theorem: there exist constants
$\varepsilon_0>0$ and $d_0\ge1$ and a polynomial-time reduction
from 3SAT to a width-at-most-three CNF formula with a satisfiable
YES case and a NO case in which every assignment violates at least
an $\varepsilon_0$ fraction of clauses, while each variable appears
in at most $d_0$ clauses
\cite{Dinur2007,Manurangsi2017rev2026,PapadimitriouYannakakis1991}.
The next two lemmas state the required normalization and disjointness
extraction explicitly.

\begin{lemma}[Normalizing to three distinct variables]
\label{lem:normal}
Under the preceding bounded-occurrence gap-3SAT source theorem, a
polynomial-time transformation produces a CNF whose clauses all have
exactly three \emph{distinct} variables, preserves perfect
completeness, retains a violated-clause fraction at least
$\varepsilon_0/8$ in the NO case, and ensures that each variable
occurs in at most $d=\max\{4d_0,8\}$ clauses.
\end{lemma}
\begin{proof}
Simplify any repeated literals in a clause and discard
identically true clauses.  The number of violated clauses of a
fixed assignment does not change under these operations; in
particular, a violated original clause cannot have been a tautology.
For a two-literal clause $a\lor b$ introduce a fresh variable $z$
and replace it by
\[
 (a\lor b\lor z),\qquad (a\lor b\lor\neg z).
\]
For a one-literal clause $a$ introduce two fresh variables $u,v$
and replace it by the four clauses
\[
 a\lor u\lor v,\qquad
 a\lor u\lor\neg v,\qquad
 a\lor\neg u\lor v,\qquad
 a\lor\neg u\lor\neg v.
\]
An empty clause, if retained rather than handled as an immediate
NO instance, can be replaced by all eight sign patterns on three
fresh variables.  An existing three-literal clause on distinct
variables is unchanged.  In each gadget, if the original clause
is true, \emph{all} new clauses can be satisfied; if it is false,
\emph{every} assignment to the fresh variables violates at least one
new clause.  These statements can be checked directly, including
for unit clauses: when $a$ is false exactly one of the four sign
patterns is false for each $(u,v)$.

There are at most eight replacement clauses per source clause,
and a source variable is copied into at most four clauses per
original occurrence.  Fresh variables occur in at most eight
clauses.  Therefore the NO fraction of violated clauses decreases
by a factor of at most eight, while the YES case stays satisfiable
and all scopes have exactly three distinct variables.  Remove
unused variables so that the resulting number of variables is at
most three times its clause count.
\end{proof}

\begin{lemma}[Extracting variable-disjoint violated constraints]
\label{lem:matching}
Suppose an exact-ternary CNF has $n$ nonisolated variables and $m$
clauses, each variable occurs in at most $d$ clauses, and every
assignment violates at least $\varepsilon m$ clauses.  Then every
assignment violates at least
\[
 \gamma n,\qquad \gamma=\frac{\varepsilon}{9d},
\]
clauses whose three-variable scopes are pairwise disjoint.
\end{lemma}
\begin{proof}
Fix an arbitrary global assignment, and consider the hypergraph
whose edges are its violated clauses.  Greedily choose a violated
clause, and delete all remaining violated clauses sharing at least
one of its three variables.  At most $3d$ clauses disappear per
step, counting the chosen clause itself.  Starting from at least
$\varepsilon m$ violated clauses, the greedy algorithm selects at
least $\varepsilon m/(3d)$ clauses on pairwise disjoint scopes.
Every variable is contained in some clause, so $n\le3m$.
Consequently $\varepsilon m/(3d)\ge\varepsilon n/(9d)$.
The chosen collection can depend on the global assignment, as
allowed by Theorem~\ref{thm:dks}.
\end{proof}

\begin{corollary}[Fixed-$\ell$ polynomial-size near-clique hardness]
\label{cor:fixedhard}
Under the standard bounded-occurrence gap-3SAT theorem, for every
\emph{fixed integer} $\ell\ge2$ there is a constant $c_\ell>0$
such that it is NP-hard to distinguish, for an explicitly given
simple graph with $N$ vertices and target size $K$, between
\begin{enumerate}
\item a YES case in which the graph contains a $K$-clique, and
\item a NO case in which every induced $K$-vertex subgraph has
edge density at most $1-c_\ell N^{-2/\ell}$.
\end{enumerate}
This is an \emph{additive near-clique promise} result.  It does not
assert constant-factor inapproximability for general Densest
$k$-Subgraph.
\end{corollary}
\begin{proof}
Apply Lemmas~\ref{lem:normal} and \ref{lem:matching} to a
bounded-occurrence gap-3SAT instance.  The resulting NO instances
satisfy the disjoint-violation promise with a fixed
$\gamma=(\varepsilon_0/8)/(9d)=\varepsilon_0/(72d)>0$.
For all sufficiently large instances, $n\ge16/\gamma$ and
$\ell\le\sqrt n$; finitely many smaller instances do not affect
NP-hardness.  By \eqref{eq:dks-sizes}, the output graph has
$K=\binom n\ell$ and
$N=2^\ell\binom n\ell=\Theta_\ell(n^\ell)$.
An adjacency test inspects at most $m$ clauses and the assignments
on at most $2\ell$ coordinates.  Explicitly generating the graph
therefore takes $O(N^2(m+\ell))$ time (up to ordinary data-structure
factors), which is polynomial in the source instance size for fixed
$\ell$.  In the YES case Lemma~\ref{lem:complete} gives a $K$-clique;
in the NO case Theorem~\ref{thm:dks} gives missing-edge fraction at
least
\[
 \frac{\gamma^2}{13600}\frac{\ell^3}{n^2}
 =\Omega_\ell(N^{-2/\ell}).
\]
Absorb the implicit constant into $c_\ell$.
\end{proof}

\begin{remark}[Growing $\ell$ and fine-grained interpretations]
\label{rem:growing-ell}
For growing $\ell$, the same explicit graph has
$N=2^\ell\binom n\ell$ and takes $O(N^2(m+\ell))$ time to
construct.  In the range $2\le\ell\le\sqrt n$,
$\log N=\Theta(\ell\log(n/\ell))$; for
$\ell=n^a$ with fixed $0<a\le1/2$, both output size and
construction time are $2^{O(n^a\log n)}$ when $m=O(n)$.
These are \emph{size and time bounds}, not hardness assumptions.
A hardness-vs-time claim under Almost-ETH requires the source
PCP blowup and the target algorithm's running time to be counted
end to end.  Our additive gap does not supersede Manurangsi's
stronger multiplicative DkS hardness theorem under that
assumption~\cite{Manurangsi2026Revision}.  The earlier ETH/Gap-ETH
formulations of the latter work were updated in its October 2026
revision; we do not rely on either assumption here.
\end{remark}

\section*{AI methodology}
OpenAI ChatGPT (GPT-6) was used substantively in the development,
reconstruction, and revision of mathematical arguments (including
entropy estimates), in drafting and editing the paper, in locating and
comparing relevant literature, and in writing and running finite
computational checks.  AI-generated referee feedback, including Google
DeepReview, was used to examine possible proof gaps and to improve
notation, exposition, and the discussion of limitations.  An earlier
AI report identified an omitted case in the terrible-set entropy
estimate, which was subsequently addressed in the written proof.

\bibliographystyle{alpha}
\bibliography{stoc_refs}
\end{document}